\documentclass[twocoonelumn,aps,showpacs,preprintnumbers,amsmath,amssymb, prl,
superscriptaddress]{revtex4}
\usepackage{amsmath}
\usepackage{graphicx}
\usepackage{bm}

\begin{document}

\title{Formation of collective excitations in quasi-one dimensional metallic nanostructures: size and density dependance.}

\author{Amy Cassidy}
\affiliation{Department of Physics and Astronomy, University of Southern
California, Los Angeles, California 90089, USA}
\author{Ilya Grigorenko}
\affiliation{Theoretical Division T-11, Center for Nonlinear
Studies, Center for Integrated Nanotechnologies, Los Alamos National
Laboratory, Los Alamos, New Mexico 87545, USA}
\author{Stephan Haas}
\affiliation{Department of Physics and Astronomy, University of Southern
California, Los Angeles, California 90089, USA}

\date{\today}

\begin{abstract}
We investigate theoretically the formation of collective excitations
in atomic scale quasi-one dimensional metallic nanostructures. The
response of the system is calculated within the linear response
theory and random phase approximation. For uniform nanostructures a
transition from quantum single particle excitations to classical
plasmon scaling is observed, depending on the system length and
electron density. We find crucial differences in the scaling
behavior for quasi-one dimensional and three-dimensional
nanostructures. The presence of an additional modulating on-site
potential is shown to localize electrons, leading to the response
function that is highly sensitive to the number of electrons at low
fillings.

\end{abstract}


\maketitle

The creation, amplification and control of plasmon excitations in
nanostructures (nanoplasmonics)  promise an extreme usefulness in
near-field scanning microscopy, single molecule detection and other
applications \cite{sers,nanorice}. The recent advances in the
fabrication and control of low-dimensional nanostructures at atomic
resolution (for example, see \cite{ho,ho1}) make possible control of
the interaction of these systems with electromagnetic radiation on a
quantum-mechanical level. For example, clusters of Au atoms,
arranged in linear chains on a NiAl(100) surface \cite{ho}, promise
to be good candidates for the purpose of local field enhancements,
as well as for other applications, including optical media with a
negative refractive index, and subwavelength focusing of
electromagnetic radiation. Thus, it becomes necessary to examine the
fundamental aspects of light-matter interaction of these systems
with electromagnetic radiation.



It is known that in small nanostructures one can observe
{\it both} collective modes and single particle excitations (for example,
see \cite{haberland,grigorenko}). Since for many situations strong local field
enhancement in nanostructures is due to collective plasmon
excitation \cite{stockman}, it becomes a fundamental question: how many
electrons is enough to create collective (plasmon) response in a
metallic nanostructure, and what is the role of the nanostructure's
geometry in the formation of the collective response?

Because of the significant inherent finite-size gaps in such
systems, it is essential to perform a detailed quantum-mechanical
analysis that identifies the relevant energy scales for the
different types of modes. In this paper we study the nature of
electronic excitations in quasi-one dimensional metallic
nanostructures, investigating the transition between single-particle
and collective response as a function of system size, electron
density and on-site potential.

In the present approach, the excitation spectrum is determined using
linear response theory within the random phase approximation (RPA).
We consider how the characteristic excitation energies scale
with the system size and electron density.

First, we find that for a fixed
number of electrons $N_{el}$, and variable size $L$ of a quasi-one dimensional system,
there is a transition from multiple single-particle excitations to a single dominant collective plasmon-like
resonance. The typical size at which the single-particle excitations converge into a single plasmon peak, we denote as the critical size $L_{cr}$. For relatively small sizes of the system the observed $L^{-2}$ scaling of the single-particle transitions can be qualitatively explained in terms of a quantum particle in a box picture. We denote this regime as "quantum".
For larger sizes of the system the plasmon frequency scales as  $L^{-1/2}$, that is
consistent with the classical plasmon resonance frequency scaling $\omega_p\propto\sqrt{n}$, where $n=N_{el}/L$ is the electron density. We denote this regime as "classical".

Second, we investigate how the critical size $L_{cr}$ depends on the
electron density in quasi-one dimensional systems. It is known
\cite{grigorenko}, that for three-dimensional nanostructures the
formation of the collective response occurs for smaller system sizes
with increasing of the electron density. This is intuitively
understandable, since for higher electron densities there are more
electrons in the nanostructure which can participate in the
collective oscillations. In our simulations we find that for a fixed
size $L$ of a quasi-one dimensional system, for lower electronic
densities the plasmon resonance is recovered, whereas for higher
densities response of the system is found to be consistent with
quantum single-particle excitations. This observation is in contrast
with the optical response of three-dimensional nanostructures. We
give a qualitative explanation of the observed effect based on the
different scaling of the Fermi velocity in quasi-one dimensional and
three-dimensional quantum systems.

Furthermore, we study formation of collective response in spatially
inhomogeneous systems. We suggest that such systems can be assembled
with the same technique \cite{ho,ho1}, but using different species
of atoms. A local potential is introduced at alternating atomic
sites, and the effect of this modulation on the energy levels is
examined. A simple model of consecutive semi-infinite wells is
proposed to analyze the excitation spectrum and system response as a
function of the electron density. Finally, we investigate the
changes in the electromagnetic response for a quantum system which
undergoes a transition from extended plane waves to localized wave
functions, as the strength of the on-site potential is varied.

 Let
us consider the response to a time-varying external electromagnetic
field  with frequency  $\omega$ in a quasi-one dimensional
nanostructured system. We assume an atomic chain of $N$ atoms and
length $L$, that is modeled by a lattice with $N$ cites with the
inter-cite distance $a$. We assume the total number of electrons
$N_{el}$ in the system, and electrons can freely move along the
chain. This assumption seems to be valid since the direct
measurement of atomic chains \cite{ho} demonstrates that the
eigenstates of electrons are in a good agreement with a model of
free electrons in an infinite well potential. The Hamiltonian for an
electron with a background potential $V(x)$ can be written as $H=
-\frac{\hbar^2}{2m^*}\frac{\partial^2}{\partial x^2} + V(x)$, where
$m^*$ is the effective electron mass.  The background potential is
either taken to be uniform or varies from site to site. The
corresponding eigenproblem is solved by numerical diagonalization.

Within the RPA approximation, the dielectric function is given by
\cite{sarma}:
\begin{equation}
\epsilon^{RPA}(q,\omega)=1-V(q)\chi_0(q,\omega),
\end{equation}
where $\chi_0$ is the retarded density-density correlation function
for pair-bubble interactions. For a system with eigenenergies $E_i$
and corresponding Fermi distribution functions, $f_i=1/(\exp(E_i/k_B
T)+1)$, the density-density correlation function in one dimension is
\begin{equation}
\chi_0(q,\omega)=\frac{1}{L}\sum_{ij}\frac{f_i-f_j}{\hbar(\omega+
i\gamma)+E_i-E_j} |M_{ij}|^2,
\end{equation}
with matrix elements $ M_{ij} = \langle i |e^{iqx}|j\rangle $
between eigenstates $i$ and $j$. $\gamma$ is a small level
broadening constant. The Fourier transform of the Coulomb potential
in one dimension  is given by $V(q) =4\pi e^2 \text{ln}(qa)$ \cite{sarma}, where
the lattice spacing $a$ is used as a cut-off for the low-$q$
divergence. In the following calculations we focus on the
long-wavelength limit $ q\rightarrow 0 $.

From the dielectric response, the loss function, ${\cal{L}}(q,\omega)$ can be obtained via
\begin{equation}
{\cal{L}}(q,\omega)=\text{Im} \; \epsilon^{-1}(q,\omega).
\end{equation}
Collective (plasmon) excitation  appears as a large spike in the
loss function, which occur when the real part of $\epsilon$ vanishes
and the imaginary part of $\epsilon$ is sufficiently small.

Let us now focus on the dependence of the characteristic resonance
frequencies $\omega_i$ on the size of the system, specifically
looking for a possible transition from quantum to classical scaling.
The excitation energies can be determined by the zeros in the
dielectric function, and depend on the difference between the
eigenenergies of occupied and unoccupied states. Let us give an
analysis of the two limiting cases: for small system sizes and large
system sizes, correspondingly.

For small system sizes, one can describe the system as a quantum mechanical infinite square well of the length $L$.
The eigeneneries are expected to scale as $E_l=l^2\pi^2 t a^2 L^{-2}$, where
$l=1,2,...$ is an integer quantum number, and
$t=\frac{\hbar^2}{2m^*a^2}$ is the characteristic energy scale in
the system. In the region dominated by the single electron transitions,
the multiple poles in ${\cal{L}}(q,\omega)$ are hence expected to scale as
$L^{-2}$ in the quantum regime. In contrast, the excitation spectrum
for a quasi-one dimensional system in the limit of $L\to \infty$ and
in the long-wave-length limit $q\to0$, is known to be dominated by a
collective response at the plasma frequency \cite{sarma}:
\begin{equation}
\label{plasmon} \omega_p \approx q a \omega_0|\ln(qa)|^{1/2}+O(q^2),
\end {equation}
with $\omega_0=\sqrt{2 n e^2/m^*a^2}$, where $n$ is the electron
density. Keeping the number of electrons in the system fixed, the
electron density scales as $n\propto L^{-1}$, and
the plasmon energy is expected to scale as $\omega_p \propto L^{-1/2}$. This
scaling behavior is also valid for three-dimensional
nanostructures, assuming scaling of only one dimension of a
nanostructure, while keeping the other two dimensions fixed.

To summarize, from our analysis, one can predict two different scaling
regimes for frequencies of the most pronounced excitations in the
system $\omega_i$, with the changes of the system size, while the
number of electrons $N_{el}$ is fixed. The first scaling regime is
$\omega_i\propto L^{-2}$, in this regime most excitations have
single-particle nature. We called this regime "quantum". In the second
scaling regime the single particle poles merge into one plasmon resonance $\omega_p$, which scales as $\omega_p\propto L^{-1/2}$. We denote this
regime as "classical".

\begin{figure}
\includegraphics[scale=0.6]{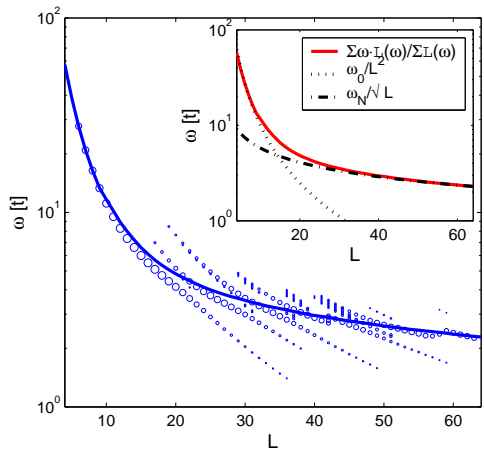}
\caption{\label{fig1}Low energy poles in ${\cal{L}}(q,\omega)$ as a
function of system size, $L$ (in units of the lattice spacing, $a$),
for a fixed number of electrons $N_{el}=50$. The solid line is a weighted
average of ${\cal{L}}(q,\omega)$ in frequency space. $\gamma$=0.05
t, q=0.05 $a^{-1}$. In the inset is shown the transition from
quantum $L^{-2}$ to classical $L^{-1/2}$ scaling.}
\end{figure}

In Fig.~1 the poles of the dielectric response function  are plotted
as a function of the system size for a fixed number of electrons of $N_{el}=50$,
considering chain lengths with $4$ to $64$ sites.  Each
circle represents a peak in the loss function, with the diameter of
the circle proportional to its strength. The solid line is the
weighted average over all the poles, given by
\begin{equation}
\label{aver} \bar{\omega}=\frac{\sum_i \omega_i \times {\cal{L}}(q,
\omega_i)} {\sum_i{\cal{L}}(q,\omega_i)}.
\end{equation}
As the length of the chain is increased, the finite-size energy spacings
decrease, and the poles eventually merge into a single plasmon feature
\cite{grigorenko}.
In the inset, the weighted loss function is fitted by the
 quantum ($\propto L^{-2}$)
and classical ($\propto L^{-1/2}$) scaling forms. From the figure it
is evident that there is a smooth transition from a quantum regime
for small atomic chains to a classical regime for longer chains. The
intersection of the two fits yields a critical length scale,
$L_{cr}$ that depends on the electron density, the effective mass
and the atomic spacing. For the parameters chosen for Fig.~1,
$L_{cr}\approx 14 a$. Additionally, we check that the plasmon
resonance given by Eq.(\ref{plasmon}) is recovered in the limit of
long chains.

Now, let us investigate how the critical size $L_{cr}$ depends on
the electron density in the system. In Fig.~1 the number of
electrons has been kept fixed so that the electron density $n$ and
the dominant plasmon frequency $\omega_p$ decrease with increasing
of the system size. It is difficult to reproduce this situation with
differently doped semiconductor nanorods\cite{rods}, and impossible
with real atoms. In order to model realistic atom chains with
diffract number of atoms \cite{ho} we keep the electron density
$n=N_{el}/L$ constant and change the system size $L$.

\begin{figure}
\includegraphics[scale=0.6]{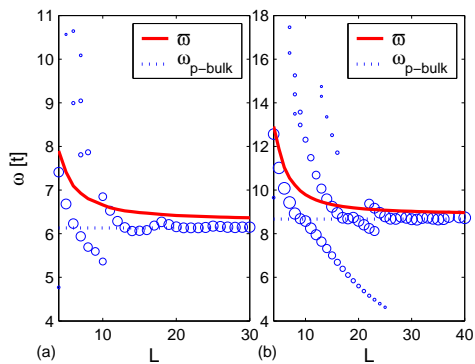}
\caption{\label{fig2} Low energy poles in ${\cal{L}}(q,\omega)$ for
constant electron density, $n$. The dotted line is the plasmon
resonance $L\to\infty$. $\gamma$ = 0.05 t. (a) $n$ = 1 electron per
site, $\omega_{p}=6.1$ t (b) $n$ = 2 electrons per site,
$\omega_{p}=8.7$ t. }
\end{figure}

In Figs.~2(a) and (b), the loss function poles are plotted as a
function of the system size for densities of $1$ and $2$ electron(s)
per site. For a fixed electron density, the single-particle
excitation frequencies are expected to scale in the quantum regime
(small $L$) as $\omega_i \propto((N_{el}+i)^2-N_{el}^2)
L^{-2}/\hbar\propto L^{-1}$, where $N_{el}=n L$, and then approach a
constant value $\omega_p$ in the bulk ($L\to\infty$). The solid line
in Fig.~2(a) represents the average frequency given by
Eq.(\ref{aver}), and the dotted line represents the plasmon
resonance, which occurs at $6.1$t for a filling of one electron per
site. For this case the critical length is determined by the length
for which the dominant pole converges to within $1\%$ of its
$L\to\infty$ value. Using this criterion, the critical length is
$L_{cr}\approx17 a$ for a filling of $1$ electron per site. In
Fig.~2(b), the plasmon resonance is $8.3$t, and the critical length
is $L_{cr}\approx 24 a$ for a filling of $2$ electron per site. For
size $L=20a$ illustrated by Fig. 2(b) one has twice electrons in the
system than in Fig.~1(a), but surprisingly, there is no established
collective response. This example illustrates the crucial difference
in response of quasi one dimensional and three-dimensional
nanostructures. For quasi-one dimensional systems the critical
length $L_{cr}$ increases with the electron density $n$, whereas it
is expected to decrease with density in three dimensions. Below we
present our analysis of the anomalous behavior of the critical
length $L_{cr}$.

The characteristic distance for
collisionless plasma is the distance  when the averaging of the
oscillating field for a moving particle happens \cite{landau}. For degenerate
electron gas it is the distance at which an electron travels at
 Fermi velocity $v_F$ during one period of the collective
field oscillation $L_{cr}\approx 2\pi v_F/\omega_p$. At the same time, the
Landau damping, which is closely connected with the spatial
dispersion of plasma, occurs also for a critical  wave vector  $k_{cr}=2\pi/L_{cr}\approx\omega_p v_F$.
One can alternatively formulate the criterion, that the collective excitations, i.e.
plasmons, do not damp for the phase velocities in plasma $v_{ph}=\omega_p/k_{cr}$ much larger than the
typical single particle velocity $v_F$ \cite{landau}. Since $\omega_p\propto\sqrt{n}$ and the Fermi velocity scales
differently with electron density $n$ in one- and three- dimensions, the
critical length depends on the density of electrons as $L_{cr}\sim
n^{1/2} $ in quasi-one dimensional ($L_{cr}\sim n^{-1/6}$ in
three-dimensional cases), consistent with the results found in
Figs.~2(a) and (b).

We suggest to check our prediction experimentally by measuring
response of atomic chains made of mono- (such as potassium or sodium) and multivalent atoms, for
example,  aluminium. Based on our analysis, formation of
collective (plasmon) response should be observed for {\it shorter} chains
made of {\it monovalent} atoms and for {\it longer} chains made of {\it multivalent} atoms.
On the contrary, in the case of three-dimensional nanostructures,
like spherical clusters, the formation of plasmon collective mode will be
observed for smaller radius of clusters made of multivalent atoms.

 Let us now consider inhomogeneous quasi-one dimensional systems made of different species of atoms.
Alternatively, one can consider a chain of coupled artificial atoms made of semiconductor quantum dots
with controlled number of electrons in the system.
Model such systems we assume an on-site potential
that takes on alternating values between consecutive atomic sites.
Starting with a small system to understand the underlying
delocalization effects, the wave functions of the low-energy
eigenstates in a double well are plotted in Fig.~3. These illustrate
the formation of bonding and anti-bonding combinations, which in the
thermodynamic limit of a multi-well chain merge into degenerate
energy bands. For clarity, the eigenfunctions in this figure are
offset by their energy gap with the ground state, $\epsilon_i -
\epsilon_0$. The eigenenergies of lowest four states, i.e. bound
states within the well, are doubly degenerate and well separated
from the higher states, reminiscent of the spectrum of an atom. As
the eigenenergies approach the barrier height, the spacing between
energies decreases and a transition occurs from localized to
delocalized states, as observed in $\psi_4 - \psi_6$.

\begin{figure}
\includegraphics[scale=0.6]{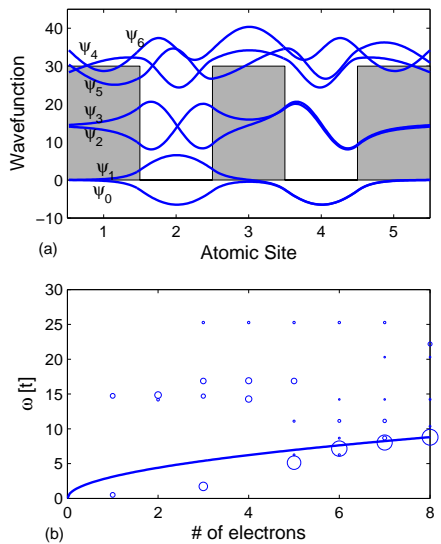}
\caption{\label{fig3}(a) Electronic wave functions of a double well
with on- site potential, $V_{os} = 30$ t, offset by $\epsilon_i -
\epsilon_0$. The on-site potential is shown by the shaded region;
(b) excitation energies in the dielectric response function for this
potential. The size of the circles represents the corresponding
oscillator strength.}
\end{figure}

For sufficiently large alternating on-site potentials and low
electron densities, the dominant low-energy excitations of larger
chains can be approximated by consecutive quasi-infinite wells,
connected via perturbative tunneling matrix elements, $\Gamma$. To
illustrate this point, consider first the case of two wells.  Using
leading-order perturbation theory, the eigenstates of this system
are $ \epsilon^{\prime}_{i\pm} = \epsilon_i \pm \Gamma$, where
$\epsilon_i$ are the eigenenergies of isolated infinite wells. This
is a good approximation as long as $\Gamma << |\epsilon_i -
\epsilon_j|$. Then, for even number of electrons $N_{el}$ , the dominant transitions
occur between the energy levels within the individual wells
$\approx$ $\epsilon_i$ and $\epsilon_{i+1}$. However, for odd number of electrons $N_{el}$,
transitions are also allowed between bonding and anti-bonding states $\epsilon^{\prime}_{i\pm} $
and $\epsilon^{\prime}_{i\mp}$.

In Fig.~3(b) the corresponding poles of the
dielectric loss function are shown for increasing number of electrons $N_{el}$ in the system.
For odd numbers of electrons at low densities, the lowest excitation
corresponds to the transition between $\psi_0$ and $\psi_1$ for one
electron and between states $\psi_2$ and $\psi_4$ for three
electrons. In contrast, the higher-energy excitations close to
$\omega = 15t$ correspond to transitions between energy bands, and
are the lowest available excitation for commensurate fillings. With
increasing number of electrons, a single low-energy pole starts to
dominate the excitation spectrum. This excitation is a precursor of
the collective plasmon mode, and it scales as $\sqrt{N_{el}}$, as
indicated by the solid line.

The effects discussed for the two-well case are amplified by
increasing the number of wells. For example, the near degeneracy of
the low-energy states increases with the number of wells, such that
for $N$ identical wells, there will be $N$ states with nearly degenerate
energies. For commensurate fillings, the excitation spectrum is then
dominated by transitions between the lowest two energy bands. As
more electrons are added, intra-band transitions, which occur on a
scale of $\Gamma$, become more dominant.

\begin{figure}
\includegraphics[scale=0.6]{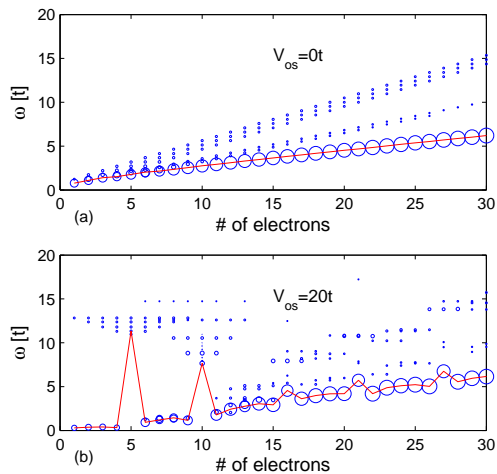}
\caption{\label{fig4} Poles in the loss function of a chain with
$N=11$ \text{sites},  $\gamma=0.05\text{ t, q}= 0.05
\text{a}^{-1}$. (a) Uniform on-site potential$ = 0$ t; (b)
alternating on-site potential = $0$ or $20$t. The thin line connects
the loss function peak with maximum intensity.}
\end{figure}

In Fig.~4, the poles in the loss function of an 11-site chain are
compared for a uniform on-site potential (a) and an alternating
potential (b). Each circle represents a peak in the loss function,
and the solid lines connect the poles with the maximum loss. In the
uniform case, Fig.~4(a), there are several energy bands that
increase {\it linearly} with increasing number of electrons, as the
difference in consecutive eigenenergies $E_{i+1}-E_{i}\propto (i+1)^2-i^2$ depends linearly on the
quantum number $i$. As more electrons are added and lower states are
filled, transitions occur between eigenstates with higher energies.

In Fig.~4(b), the low-energy poles in the dielectric loss function
are shown for a the case of an alternating on-site potential,
leading to $5$ wells of depth $20 t$. At low numbers of electrons,
the peak absorption energy increases dramatically whenever the
number of electrons is a {\it multiple} of the number of wells,
making an analog of a "closed shell". For less than five electrons,
the loss function is dominated by transitions between lowest energy
states associated with each well.  Between five and ten electrons,
the dominant transitions are between the second energy state
associated with each well. For five electrons the lowest set of
eigenstates (shell) is occupied and the dominant transition is
between the first and second energy of each well. The pattern
repeats again for ten electrons, which corresponds to two electrons
per well. Above ten electrons, the highest occupied energy level is
comparable to the well height so that the electrons are no longer
strongly localized in the wells and the excitation energies approach
the energies for the case with the uniform potential. This example
illustrates the existence of magic numbers in arrays of alternated
local potentials which are expected to be most pronounced at the
small number of electrons limit.

In conclusion, we have calculated the dielectric response function of
quasi-one dimensional nanostructures within the random phase
approximation. Using this approach, we have demonstrated that there
is a transition from classical to quantum scaling as a function of
the system size and electron density. We determine the critical $L_{cr}$ length at which collective
response is established. We find that $L_{cr}\propto n^{1/2}$ in quasi-one dimensional nanostructures,
opposite to $L_{cr}\propto n^{-1/6}$ for three dimensional nanostructures.
 We also suggested an experiment to
check our theoretical predictions. For inhomogeneous nanostructures,
modeled by alternating on-site potentials, electronic bands are
formed via tunneling between adjacent potential wells. Depending on the number of electrons,
intra- or inter-band transitions dominate the dielectric
loss function, leading to a characteristic sequence of magic
fillings. Verification of these model predictions by more
sophisticated {\it ab-initio} calculations and by luminescence
experiments, e.g. in (Al$_x$,Ga$_{1-x}$)As layered heterostructures,
are anticipated.

We are grateful to A.F.J. Levi for useful discussions and
acknowledge the Center of Integrated Nanotechnology at Los Alamos
and Sandia National Laboratories for its support.
This work was carried out under the auspices of the National Nuclear
Security Administration of the U.S. Department of Energy at Los
Alamos National Laboratory under Contract No. DE-AC52-06NA25396.

\end{document}